\documentstyle[12pt]{article}
\begin{document}
%
\newcommand{\br}{\begin{eqnarray}}
\newcommand{\er}{\end{eqnarray}}
\newcommand{\ket}[1]{\mid{#1}\rangle}
\newcommand{\go}{\longrightarrow}
\newcommand{\gsim}{\stackrel{\sim}{>}}
\newcommand{\M}{\cal{M}}
%
\begin{center}
{\bf Double beta transition mechanism }\\
\vspace{3mm}
{\small{F. Krmpoti\'{c}}\\
Departamento de F\'\i sica, Facultad de Ciencias Exactas\\
Universidad Nacional  de La Plata, C. C. 67, 1900 La Plata, Argentina}
\end{center}
\date{}
\begin{abstract}
After briefly reviewing $\beta \beta$ decay as a test of the neutrino mass, I
examine the nuclear structure involved in this process.
Simple formulas (\`{a} la Pad\'{e}) are designed for the transition amplitudes
and the general behavior of  $\beta \beta$ decay amplitudes in the
quasiparticle random phase approximation are discussed.
Results of a calculation for $^{48}Ca$, $^{76}Ge $, $^{82}Se$, $^{100}Mo$,
$^{128}Te$ and $^{130}Te$ nuclei are presented, in which the particle-particle
interaction strengths have been fixed by invoking the partial restoration of
the isospin and Wigner SU(4) symmetries.
An upper limit of  $\;<m_{\nu}>\approx\; 1$ eV is obtained for the effective
neutrino mass.
\end{abstract}

{\it PACS numbers}: 23.40.Hc, 21.10.Re, 21.60.Jz

\bigskip

{\bf 1. Introduction}

The double beta ($\beta \beta$) decay is a nice example of the
interrelation between the Particle Physics and the Nuclear Physics:
{\it we can get information on the properties of the neutrino and the weak
interaction from the $\beta \beta$ decay only if we know who to deal we the
nuclear structure involved in the process}. There are already several
well-known reviews on the neutrino physics \cite{Reviews,Tom91} and therefore,
after a brief historical overview,
I will limit this talk to the nuclear facet of the problem. More precisely,
some recent developments performed by our group will be summarized.

Because of the pairing force there are approximately 50 nuclear
systems in which the odd-odd isobar, within the isobaric triplet $(A,Z)$,
$(A,Z+1)$, $(A,Z+2)$, has a higher mass than its neighbors. Within such a
scenario the single $\beta$ decay, is energetically forbidden and the
initial nucleus can disintegrate only via the $\beta \beta$ decay. This
is a second-order weak interaction process, similar to electromagnetic
processes such as the atomic Raman scattering and the nuclear $\gamma \gamma$
decay \cite{Kra87}.

The modes by which $\beta \beta$ decay can take place are connected with the
neutrino ($\nu$) - antineutrino ($\tilde{\nu}$) distinction. If
$\nu$ and $\tilde{\nu}$ are defined by the transitions:
\br
&&n \go p+e^-+\tilde{\nu}\\
\nonumber
&&\nu+n\go p+e^-,
\label{1}
\er
the decay $(A,Z) \go (A,Z+2)$ can occur by successive $\beta$ decays:
\br
(A,Z)&\go &(A,Z+1)+e^-+\tilde{\nu}
\nonumber\\
&\go &(A,Z+2)+2e^-+2\tilde{\nu},
\label{2}
\er
passing through the intermediate virtual states of the $(A,Z+1)$ nucleus.

Yet the neutrino is the only fermion in lacking a additionally
conserved quantum number that differences between $\nu$ and $\tilde{\nu}$.
Thus it is possible that the neutrino is a Majorana particle, i.e., equal to
its own
antiparticle (\`{a} la $\pi^0$).
\footnote {A Dirac particle can be viewed as a combination of two Majorana
particles
with equal mass and opposite CP properties and their contribution to the
$\beta \beta_{0\nu}$ decay cancel.}
When $\tilde{\nu}=\nu $ the process
\br
(A,Z)&\go &(A,Z+1)+e^-+\tilde{\nu}
\equiv (A,Z+1)+e^-+\nu
\nonumber\\
&\go &(A,Z+2)+2e^-
\label{3}
\er
is also allowed.
In absence of the helicity suppression (as would be natural before the
observation of parity violation) this neutrinoless ($\beta \beta_{0\nu})$
mode is favoured by phase space over the two-neutrino ($\beta \beta_{2\nu})$
mode by a factor of $10^7-10^9$: $T_{0\nu}\sim (10^{13}-10^{15})$ y while
$T_{2\nu}\sim (10^{20}-10^{24})$ y. Several searches for the $\beta \beta$
decay has been made by the early 1950s with the result that $T\gsim 10^{17}$.
This seemed to point that $\tilde{\nu}\neq \nu$ and prompted the introduction
of the lepton number $L$ to distinguish $\nu$ from $\tilde{\nu}$: $L=+1$ was
attributed to $e^-$ and $\nu$ and $L=-1$ to $e^+$ and $\tilde{\nu}$.
The assumption that the additive lepton number is conserved then allows
the $\beta \beta_{2\nu}$ decay but prohibits the $\beta \beta_{0\nu}$ one,
for which $\Delta L=2$.

But with the discovery in 1957 that the parity is not conserved for the weak
interaction it was realized that the Majorana/Dirac character of the neutrino
was still in question. If
\br
&&n \go p+e^-+\nu_{RH}\\
\nonumber
&&\nu_{LH}+n\go p+e^-
\nonumber
\er
then the second process in (\ref{3}) is forbidden because the right handed
neutrino has the wrong helicity to be reabsorbed. Therefore the $\gamma_5$-
invariance of the weak interaction could account for no
$\beta \beta_{0\nu}$ decay, regardless of the Dirac or Majorana nature of the
neutrino. Otherwise, this decay can be observed only when the lepton
number is not conserved and the neutrino is a massive Majorana particle.
\footnote{We assume for simplicity that weak interactions with right-handed
currents do not play an essential role in the neutrinoless mode.}
This event discouraged experimental searches for a long time, but
with the development of modern gauge theories the situation began to change.
In fact, there are many reasons for the renaissance of interest in
$\beta \beta$-decay over the past decade. The most important one is that,
if there is any new physics beyond the standard $SU(2)_L\times U(1)$
gauge model of electroweak interactions, the $\beta \beta_{0\nu}$ decay will
play a crucial role in shaping the ultimate theory.
Moreover, no solid theoretical principle precludes neutrinos from having
mass and the most attractive extensions of the standard model require neutrinos
to be massive. The theory is neither capable of predicting the scale of
neutrino masses any better than it can fix the masses of quarks and
charged leptons.

Since the half-lives can be cast in the simple form
\[
T_{2\nu}^{-1}={\cal G}_{2\nu}{\cal M}_{2\nu}^2,
\hspace{2cm}
T_{0\nu}^{-1}={\cal G}_{0\nu}{\cal M}_{0\nu}^2 <m_{\nu}>^2,
\]
(where ${\cal G}'s$ and ${\cal M}'s$ are, respectively, the phase space
factors and nuclear matrix elements and $<m_{\nu}>$ is the effective neutrino
mass) it is clear then that we shall not understand the $\beta \beta_{0\nu}$
decay unless we understand the $\beta \beta_{2\nu}$
decay. The last is one of the slowest process observed so far in nature and
offers a unique opportunity for testing the nuclear physics techniques for
half-lives $\gsim 10^{20}$ y.  Thus, the comprehension of the $\beta \beta$
transition mechanism cannot but help advance knowledge of physics in general.

It is worth noting that more than 30 $\beta\beta$ decay experiments are
underway or in stages of planning and construction. Until now positive
evidence of the $\beta \beta_{2\nu}$  decay mode has been found for
$^{76}Ge$, $^{82}Se$, $^{100}Mo$, $^{128}Te$, $^{130}Te$ and $^{238}U$.
Yet, despite the colossal experimental progress the
neutrinoless, lepton violating decay, if it exists, has escaped detection
until now.
\footnote{ The $\beta \beta_{0\nu}$  sensitivities have changed from
$\sim 5\times 10^{15}$ in 1948 to $\sim 5\times 10^{23}$  in 1987.}

The $\beta \beta$ decays occur in medium-mass nuclei that are rather far
from closed shells, and we all know that shell-model calculations are
practical only when the number of valence nucleons is relatively small.
Therefore, at the present time, the nuclear structure method most widely used
is the quasiparticle random phase approximation (QRPA). Within this model the
$\beta \beta$-decay amplitudes are very sensitive to the interaction parameter
in the particle-particle (PP) channel, usually denoted by $g^{pp}$.
It is still more interesting that, close to the expected value for $g^{pp}$,
the $\beta \beta$ matrix elements go to zero. But when a physical quantity
has a zero (or near zero) a conservative law should, very likely, be at its
origin.
Thus, resorting to a toy model, I will first discuss the general
behavior of the nuclear matrix elements \cite{Krm92,Krm93}. Later, I will
show that they have zeros because of the restoration of the isospin and
Wigner $SU(4)$ symmetry \cite{Hir90,Krm94}. This is not surprising since the
Fermi (F) and Gamow-Teller (GT) operators $\tau_{\pm}$ and $\sigma \tau_{\pm}$,
relevant in the $\beta \beta$ decay, are infinitesimal generators of $SU(2)$
and $SU(4)$, respectively. Finally, we use the concept of restoration
of these symmetries to fix the PP interaction strengths and to estimate the
$\beta \beta$ matrix elements. It can be argued that, the
$SU(4)$ symmetry is badly broken in heavier nuclei and that therefore
our recipe is quite arbitrary. I will show, yet, that the residual
interaction is capable to overcome most of the $SU(4)$ breaking
caused by the spin-orbit splitting.

{\bf 2. General behavior of ${\cal M}_{2\nu}$ and ${\cal M}_{0\nu}$ in the
QRPA}

Independently of the nucleus that decays, of the residual interaction
that is used, and of the configuration space that is employed, the
$\beta \beta$-moments as a function of $g^{pp}$ always exhibit the
following features:

(i) The $2\nu$ moments have first a zero and latter on a pole at which the
QRPA collapses.

(ii) The  zeros and  poles of ${\cal M}_{0\nu}$ for the virtual
states with spin and parity $J^{\pi}=1^{+}$ are strongly correlated with the
zeros and poles of ${\cal M}_{2\nu}$.

(iii) The  total $\beta \beta_{0\nu}$ moments
also possess zeros but at significantly larger values of $g^{pp}$.

The behaviour of the $\beta \beta$ moments for several nuclei are illustrated
in Fig.\ \ref{fig1}. These results have been obtained with a $\delta $ force,
using standard parametrization \cite{Krm94}. Instead of the parameter $g^{pp}$,
I use here the ratio  between the triplet and singlet coupling strengths in
the PP channel, i.e., $t=v_t^{pp}/v_s^{pp}$. Calculations with finite range
interactions yield similar results \cite{Tom91}.

I will resort now to the single mode model  (SMM) description
\cite{Krm92,Krm93} of the $\beta \beta$-decays in the
$^{48}Ca \rightarrow\, ^{48}Ti$ and $^{100}Mo \rightarrow\, ^{100}Ru$ systems.
This is the simplest version of the QRPA with only one intermediate state for
each $J^{\pi}$. It allows to express the moments \`{a} la Alaga
\cite{Ala71}, i.e., as the unperturbed matrix elements ${\cal M}_{2\nu}^0$ and
${\cal M}_{0\nu}^0 (J^{+})$ multiplied by the effective charges:
\begin{eqnarray}
{\cal M}_{2\nu}&=& {\cal M}_{2\nu}^0  \left(\frac{\omega^{0}}
{{\omega}_{1^{+}}}
\right)^{2}\, \left(1+\,\frac{{\it G}(1^{+})} {{\omega}^{0}}\right),
\label{5}\\
{\cal M}_{0\nu} (J^{+})&=& {\cal M}_{0\nu}^0 (J^{+})\,
 \frac{\omega^{0}} {{\omega}_{J^{+}}}\,
\left(1+\,\frac{{\it G}(J^{+})} {{\omega}^{0}}\right),
\label{6}
\end{eqnarray}
Here  ${\it G}(J^{+})\equiv{\it G}(pn,pn;J^{+})$ are the PP matrix elements
(proportional to $t$ (or to $g^{pp}$)), $\omega^{0}$ is the unperturbed energy,
and ${\omega}_{J^+}$ are the perturbed energies. It will be assumed here that
the isospin symmetry is strictly conserved, in which case (as it will be seen
latter on) ${\cal M}_{2\nu}(0^{+})= {\cal M}_{0\nu}(0^{+})\equiv 0$.
When the pairing factors are estimated in the usual manner, one gets
\begin{equation}
{\omega}={\omega}^0\sqrt{1+{\it F}(34+9{\it F}/{\omega}^0)/25{\omega}^0+
16{\it G}(1+{\it F}/{\omega}^0)/25{\omega}^0},
\label{7}
\end{equation}
for the single pair configurations $[0f_{7/2}(n)0f_{7/2}(p)]_{J^+}$ in
$^{48}Ca$ and
\begin{equation}
{\omega}={\omega}^0\sqrt{1+4{\it F}(45+{\it F}/{\omega}^0)/225{\omega}^0+
{\it G}(270+172{\it F}/{\omega}^0+49{\it G}/{\omega}^0)/225{\omega}^0},
\label{8}
\end{equation}
for $[0g_{7/2}(n)0g_{9/2}(p)]_{J^+}$ in $^{100}Mo$. Therefore, while the
numerators in Eqs. (\ref{5}) and (\ref{6}) depend only on the PP matrix
elements, their denominators depend on the particle-hole (PH) matrix elements
${\it F}(J^{+})\equiv{\it F}(pn,pn;J^{+})$, as well.
The numbers in the last two equations arise from the pairing factors.
As illustrated in Fig.\ \ref{fig2}, the SMM is a fair first-order
approximation for the $\beta \beta_{2\nu}$ decays in $^{48}Ca$ and
$^{100}Mo$ nuclei.

The role played by the ground state correlations (GSC)
in building up Eqs. (\ref{5}) and (\ref{6}) can be summarized as follows:\\
(a) The numerator, i.e., the factor $(1+{\it G}/{\omega}^0)$, comes from the
interference between the forward and backward going contributions.
These contribute coherently in the PP channel and totally out of phase in
the PH channel.\\
(b) The ${\it G}^2$ and ${\it F}^2$ terms in the denominator are
very strongly quenched by the GSC, while the  ${\it GF}$ term is enhanced
by the same effect. In particular, for $^{48}Ca$ the term quadratic in
${\it G}$  does not contribute at all.\\
It can be stated therefore that, within the SMM and because of the GSC,
the ${\cal M}_{2\nu}$ matrix element is mainly a bilinear function of
${\it G}(1^{+})$.
Besides, it passes through zero at ${\it G}(1^{+})=-{\omega}^0$ and  has a
pole when ${\omega}_{1^{+}}=0$.
Similarly, all ${\cal M}_{0\nu}(J^{+})$ moments turn out to be quotients of
a linear function of ${\it G}(J^{+})$ and the square root of another linear
function of ${\it G}(J^{+})$. Both the zero and the pole of
${\cal M}_{0\nu}(1^{+})$ matrix element coincide with those of the $2\nu$
moment. Besides, as the magnitudes of ${\it G}(J)$ and ${\it F}(J)$ decrease
fairly rapidly with J (see Table \ref{tab0}),
the quenching effect, induced by the PP interaction, mainly
concerns the allowed $0\nu$ moment. For higher order multipoles it
could be reasonable to expand the denominator in Eq. (\ref{6}) in powers of
${\it G}(J^{+})/{\omega}^0$ and to keep only the linear term. This term
strongly cancels with a similar term in the numerator and the net result is a
weak linear dependence of the  ${\cal M}_{0\nu}(J^{+}\neq 1^{+})$ moments on
the PP strength. Obviously, for the last approximation to be valid, the
parameter  $t$ (or $g^{pp}$) has to be small enough to keep ${\omega}_{1^{+}}$
real.  Briefly, the SMM  can account for all four points raised
above, and leads to the following approximations
\begin{equation}
{\cal M}_{2\nu}\cong {\cal M}_{2\nu}(t=0)\frac{1-t/t_0}{1-t/t_1},
\label{9}
\end{equation}
and
\begin{eqnarray}
{\cal M}_{0\nu}&\cong& {\cal M}_{0\nu}(J^{\pi}= 1^{+};t=0)
\frac{1-t/t_0}{\sqrt{1-t/t_1}}\nonumber\\
&+&{\cal M}_{0\nu}(J^{\pi}\neq 1^{+};t=0)(1-t/t_2),
\label{10}
\end{eqnarray}
where $t_1 \geq t_0$ and $t_2 \gg t_1$, and the condition $t\leq t_1$ is
fulfilled.
It is self evident that these formulae do not depend on the type of residual
interaction, and that analogous expressions are obtained
when the parameter $g^{pp}$ is used (with $g^{pp}$'s for t's).

The common behavior of the $\beta \beta$ moments for all nuclei, together
with the similarity between the SMM and the full calculations for
$^{48}Ca$ and $^{100}Mo$ (shown in Figs. \ref{fig1} and \ref{fig2},
respectively), suggests to go a step further and try  to express the
exact calculations within the framework of Eqs. (\ref{9}) and (\ref{10}).
At a first glance this seems a difficult task, because:
(i) the SMM does not include the effect of the spin-orbit splitting, which
plays a very important role in the $\beta \beta$-decay through the dynamical
breaking of the SU(4) symmetry, and (ii) the full calculations involve a
rather large configuration space (of the order of 50 basis vectors).

The parameters $t_0$, $t_1$, and $t_2$ that fit the  $\beta \beta$ moments
displayed in Fig.\ \ref{fig1} are listed in Table \ref{tab1},
together with moments ${\cal M}_{2\nu}$, ${\cal M}_{0\nu}(J^{\pi}= 1^{+})$,
and ${\cal M}_{0\nu}(J^{+}\neq 1^{+})$ for $t=0$.
The reliability of formulae (\ref{9}) and (\ref{10}) is surprising,
to the extent that it is not possible  to distinguish visually  the exact
curves from the fitted ones. It is worth noting that this situation
persists even within the number projected QRPA \cite{Krm93a}.
Why the exact calculations can be accounted for by Eqs. (\ref{9}) and
(\ref{10})?
I do not know a fully convinced answer. Yet, let me note that for a $n$
dimensional configuration space, ${\cal M}_{2\nu}$ can always be expressed by
the ratio of the polynomials of degrees $2n-1$ and $2n$ in ${\it G}(1^{+})$
 \cite{Hir90}, i.e.,
\begin{equation}
{\cal M}_{2\nu}\cong {\cal M}_{2\nu}(t=0)
\frac{1-t/t_0^{(1)}-t^2/t_0^{(2)}- \cdot \cdot -t^{2n-1}/t_0^{(2n-1)}}
{1-t/t_1^{(1)}-t^2/t_1^{(2)}-\cdot \cdot -t^{2n}/t_0^{(2n)}},
\label{11}
\end{equation}
Thus the above results seem to indicate that cancellations of the type (a) and
(b) are likely to be operative to all orders, and that linear terms in
${\it G}(1^{+})$ are again the dominant ones. General expressions for
${\cal M}_{0\nu}$, analogous to (\ref{11}), are not known, but some
cancellation must be taking place in these as well.

{\bf 3. Restoration of the isospin and SU(4) symmetries}

An important question in the QRPA calculations is, how to fix $g^{pp}$ or $t$?
Several attempts have been made to calibrate $g^{pp}$ using the experimental
data
for  individual GT positron decays. The weak point of this
procedure is that the distribution of the $\beta^{+}$ strength among
low-lying states in odd-odd nuclei is certainly affected by the
charge-conserving vibrations, which are not included in the QRPA. For example,
the single beta transitions $^{100}Tc\rightarrow{^{100}Mo}$ and
$^{100}Tc \rightarrow{^{100}Ru}$ have been discussed recently in the standard
QRPA \cite{Gri92}, where the $^{100}Tc$ states are described as pure
pn-quasiparticle excitations,
while the suggested wave function for the ground state in $^{101}Mo$ is
(cf. ref. \cite{Sey91}) is only $\approx 35\%$ of quasiparticle nature:
\begin{eqnarray*}
\ket{1/2^+}&=&0.59\ket{s_{1/2},00}-0.57\ket{s_{1/2},20}+0.32\ket{s_{1/2},40}\\
 &-&0.26\ket{d_{5/2},22} -0.26\ket{d_{1/2},42} -0.21\ket{g_{7/2},24}
\end{eqnarray*}
(In the basis state $\ket{j,NI}$ the quasiparticle $j$ and the $N$
bosons of angular momentum $I$ are coupled to the total spin $1/2$.)

We gauge $t$ by resorting to the restoration of the Wigner
SU(4) symmetry \cite{Hir90}.  Unlike the method mentioned above, this
method involves the  total GT strength, which dependent
of  the charge-conserving vibrations only very weakly.  We are aware, however,
that the SU(4) symmetry is badly broken in medium and heavy nuclei, and
therefore before proceeding, it is necessary to specify what we mean by
reconstruction of this symmetry.

For a system with $N\neq Z$, the isospin and spin-isospin symmetries are
violated in  the mean field approximation, even if the nuclear
hamiltonian commutes with the corresponding excitation operators
$\beta^{\pm}$ (${\tau}_{\mp}$ and $\sigma {\tau}_{\mp}$).
But, we know that when
a non-dynamical violation occurs in the BCS-Hartree Fock (BCS-HF) solution,
the QRPA induced GSC can be invoked to restore the symmetry.
There are subtleties involved in the restoration mechanism: the GSC are not
put in evidence explicitly, but only implicitly via their effects on the
one-body moments $\beta^{\pm}$ between the ground state and the excited
states. Besides, for the F excitations and when the isospin
non-conserving forces are absent, a self-consistent inclusion of the GSC
leads to the following:\\
1) all the $\beta^{-}$  strength is concentrated in the collective state,
and\\
2) the $\beta^{+}$ spectrum, which in QRPA can be viewed as an extension of
the $\beta^{-}$ spectrum to negative energies, is totally quenched.\\
The self-consitency is only attained when the same $S=0$, $T=1$
interaction coupling strengths are used in the pairing and PP channels,
i.e., when $v_s^{pair}=v_s^{pp}$, and the extent to which the above
conditions are fulfilled may be taken as a measure of the isospin symmetry
restoration. In Fig. \ref{fig4} is shown the behavior F strength $\beta^+$
as a function of the parameter $s=v_s^{pp} /v_s^{pair}$.

Besides being spontaneously broken by the HF-BCS approximation, the SU(4)
symmetry is also dynamically broken by the spin-orbit field and the
supermultiplet destroying residual interactions. But, the last two effects
have a tendency to cancel each other.
In fact, within the TDA the energy differences between the GT and F
resonances can be expressed as \cite{Nak82}
\begin{equation}
E_{GT}-E_F=\left[\Delta_{{\it l}s}-\left(v_t^{ph}-v_s^{ph}\right)
\frac{N-Z}{2A}\right]\,MeV,
\label{12}
\end{equation}
where $\Delta_{{\it l}s}\approx 20A^{-1/3}$ is the mean spin-orbit splitting
and $v_s^{ph}$ and $v_t^{ph}$ are, respectively, the singlet and the triplet
coupling constants in the PH channel. As  $v_t^{ph}>v_s^{ph}$ the
residual interaction displaces the GT resonance towards the IAS with
increasing $N-Z$.
What is more, the energetics of the GT resonances
are nicely reproduced by (see Fig. \ref{fig5})
\begin{equation}
E_{GT}-E_F=\left(26 A^{-1/3}-18.5\frac{N-Z}{A}\right) MeV,
\label{13}
\end{equation}
which has the same mass and neutron excess dependence as (\ref{12}).
Briefly, the experimental data show that the SU(4) symmetry
destroyed by the mean field is partially restored by the residual
interaction.
\footnote {Neither in $^{208}Pb$, where the GT strength is located at the
energy of the IAS, the SU(4) symmetry is totally restored, as indicated by
the resonance width of $\approx 4 MeV$.}
The GSC are likely to alter Eq. (\ref{12}) very little. But, within the QRPA
the $\sigma {\tau}_{+}$ transition strength  is strongly quenched and the GT
resonance is somewhat narrowed, as compared with the TDA results.
As such the global effect of the pn residual interaction on the GT strengths
$\beta^{\pm}$ ($\sigma {\tau}_{\mp}$) is qualitatively similar to the
corresponding effect on the  F strengths $\beta^{\pm}$ (${\tau}_{\mp}$),
in the sense that the conditions 1) and 2) are
approximately fulfilled, and we say that the SU(4) symmetry is partially
restored. It seems reasonable then to assume that the maximal restoration is
achieved for the value of $t$ where the GT strength $\beta^+$ is minimum,
and this is the way how we fix the parameter $t$.

{\bf 4. Results for  $\M_{2\nu}$ and $<m_{\nu}>$}

{}From the results displayed in Table \ref{tab2}, it can be said that with
$t= t_{sym}$ the calculated $\M_{2\nu}$  moment for $^{48}Ca$ does
not contradict the experimental limit and that the $2\nu$ measurement
in $^{82}Se$ is well accounted for by the theory. On the other
hand, the calculated $2\nu$ matrix elements turn out to be too small for
$^{76}Ge$ and $^{100}Mo$ and too large for $^{128}Te$ and $^{130}Te$ (in both
the cases by a factor of $\approx 3$).
Yet, one should bear in mind that:
i) the calculated values of $\M_{2\nu}$ vary rather abruptly near $t= t_{sym}$
and therefore it is possible to account for the $\M_{2\nu}$
in all the cases with a comparatively small variation ($<10\%$) of $t$, and
ii) the minimum value of the GT $\beta^+$ strength critically depends on the
spin-orbit splitting over which we still do not have a complete control.

Besides the issue of the procedure adopted for fixing the particle-particle
strength parameter within the QRPA, there are some additional problems in
calculations of the matrix element ${\cal M}_{2\nu}$, as yet not fully
understood. They are related with the type of force, choice of the single
particle spectra, treatment of the difference between the initial and final
nuclei, etc. All these things are to some extent uncertain and therefore it
is open to question whether it is possible, at present, to obtain
a more reliable theoretical estimate for the $2\nu$ half lives that the one
reported here.

The upper limits for the effective neutrino mass $<m_{\nu}>$, obtained from
the measured  $0\nu$  half-lives and the calculated matrix elements are shown
in Table \ref{tab3}, where also are presented the results obtained by other
groups. The difference in a factor of about $2-3$ between both:
i) the results obtained by the Pasadena group and the groups of T\"{u}bingen
and Heidelberg for $^{76}Ge$ and $^{82}Se$ nuclei, and ii) the previous and
present calculations  for $^{100}Mo$, $^{128}Te$ and $^{130}Te$ nuclei,
is just a reflection of the unavoidable uncertainty of the QRPA calculations,
and it is difficult to assess which one is "better" and which is "worse".

\noindent{\bf Acknowledgments}

I am grateful to the organizers of the workshop for their kind invitation and
hospitality, and in particular to Professor Alfonso Mondragon.
%

\newpage

\bigskip

\begin{figure}
\begin{center} { \bf Figure Captions} \end{center}
\caption{ \protect \footnotesize
Calculated matrix elements ${\cal M}_{2\nu}$ (in units of $[MeV]^{-1}$),
the $0\nu$ moments for $J^{\pi}=1^{+}$ (${\cal M}_{0\nu}(J^{\pi}=1^{+})$)
and total moments ${\cal M}_{0\nu}$ as a function of the particle-particle
$S=1$, $T=0$ coupling constant t.
\label{fig1}}
\caption{ \protect \footnotesize
Exact (solid lines) and SMM (dashed lines) matrix elements
${\cal M}_{2\nu}$ (in units of $[MeV]^{-1}$), as a function of
the  coupling constant $t/t_0$. $t_0$ is the value of $t$ for
which ${\cal M}_{2\nu}$ is null.
\label{fig2}}
\caption{ \protect \footnotesize
Calculated double beta decay matrix elements ${\cal M}_{2\nu}$
(in units of $[MeV]^{-1}$) for $^{76}Ge$, as a function of $t$.
Solid and dotted curves correspond to the projected (PQRPA)
and unprojected (QRPA) results, respectively.
\label{fig3}}
\caption{ \protect \footnotesize
Fermi and Gamow-Teller transition strengths $\beta^+$ for the nuclei
$^{48}Ca$, $^{76}Ge $, $^{82}Se$, $^{100}Mo$,  $^{128}Te$ and  $^{130}Te$,
as a function of particle-particle couplings $s$ ($S=0$, $T=1$) and $t$
($S=1$, $T=0$) respectively.
\label{fig4}}
\caption{ \protect \footnotesize
Plot of $E_{GT}-E_F$ versus $(N-Z)/A$. When the experimental results overlap
(for $^{90,92}Zr$ and $^{208}Pb$) we displace them slightly with
resect to the correct value of $(N-Z)/A$ for the sake of clarity. The values
calculated by Eq. (\protect \ref{12}) are indicated by full circles.
\label{fig5}}
\end{figure}
\begin{table} \begin{center} { \bf Tables }
\caption{ \protect \footnotesize
The ${\cal M}^0_{0\nu}(J^+)$ moments  and the factors
${\it G}(J^+)/{\omega}^0$ within the single mode model for
$^{48}Ca$ and $^{100}Mo$.}
\vspace{3mm}
\begin{tabular}{ccccc}
\hline\hline
&\multicolumn{2}{c}{$^{48}Ca$}&\multicolumn{2}{c}{$^{100}Mo$}\\
J&$-{\cal M}^0_{0\nu}$&$-{\it G}(J^+)/{\omega}^0$
&$-{\cal M}^0_{0\nu}$&$-{\it G}(J^+)/{\omega}^0$\\
\hline
0&1.0159& $  s               $&        &$                                   $\\
1&1.3439& $ \frac{11}{21}t   $&  2.8316&$  \frac{20}{27}t                   $\\
2&0.1573& $ \frac{5}{21}s    $&  0.2741&$  \frac{20}{77}(t+\frac{2}{27}s)
$\\
3&0.2143& $ \frac{19}{77}t   $&  0.2086&$  \frac{130}{693}t                 $\\
4&0.0446& $ \frac{9}{77}s    $&  0.1263&$
\frac{162}{1001}(t+\frac{20}{81}s)$\\
5&0.1081& $ \frac{235}{1001}t$&  0.0585&$  \frac{124}{1287}t                $\\
6&0.0122& $ \frac{25}{429}s  $&  0.0842&$  \frac{20}{143}(t+\frac{14}{27}s)
$\\
7&0.0988& $ \frac{175}{429}t $&  0.0177&$  \frac{190}{3861}t                $\\
8&      & $                  $&  0.0925&$  \frac{490}{2431}(t+\frac{8}{9}s)
$\\
\hline\hline
\end{tabular}
\label{tab0}
\caption{ \protect \footnotesize
The coefficients $t_0$, $t_1$, and $t_2$ and the matrix elements
${\cal M}_{2\nu}$, ${\cal M}_{0\nu} (J^{\pi}= 1^{+})$, and
${\cal M}_{0\nu} (J^{\pi}\neq 1^{+})$ for $t=0$, in the parametrization of
the $2\nu$ and $0\nu$ $\beta \beta $ moments.
The matrix elements ${\cal M}_{2\nu}$ are given in units of $[MeV]^{-1}$.
The values of the PP coupling strengths, which lead to maximal
restoration of the SU(4) symmetry ($t=t_{sym}$), are shown in the last row.}
\label{tab1}
\vspace{3mm}
\begin{tabular}{c|ccccccc}
\hline\hline
&{$^{48}Ca$}&{$^{76}Ge$}&{$^{82}Se$}&{$^{100}Mo$}&{$^{128}Te$}&{$^{130}Te$}\\
\hline
$-{\cal M}_{2\nu}$ &0.173&0.308&0.321&0.451&0.381&0.331\\
$t_0$    &1.394& 1.161&1.206 &1.469 &1.265 &1.261 \\
$t_1$    &1.754& 1.680&1.691 &1.649 &2.131 &2.268\\
\hline
$-{\cal M}_{0\nu} (J^{\pi}= 1^{+})$ &1.506&4.242&4.179&5.015&4.599&4.182\\
$-{\cal M}_{0\nu} (J^{\pi}\neq 1^{+})$ &1.501&6.924&7.495&9.762&7.997&7.486\\
$t_0$     &1.227& 1.155&1.141 &1.372 &1.377 &1.407 \\
$t_1$     &1.768& 1.741&1.764 &1.711 &2.236 &2.345 \\
$t_2$     &12.82& 13.23&12.14 &6.527 &13.39 &11.08 \\
\hline
$t_{sym}$     &$\cong 1.50$&$\cong 1.25$&$\cong 1.30$&$\cong 1.50$
&$\cong 1.40$&$\cong 1.40$\\
\hline\hline
\end{tabular}
\end{center} \end{table}
\newpage
\begin{table} \begin{center}
\caption{ \protect \footnotesize
Experimental and calculated ${2\nu}$ moments for $t=t_{sym}$
(in units of $[MeV]^{-1}$).}
\vspace{3mm}
\begin{tabular}{cccccccc}
\hline\hline
&{$^{48}Ca$}& {$^{76}Ge$}& {$^{82}Se$}& {$^{100}Mo$}&
{$^{128}Te$}& {$^{130}Te$}\\
\hline
$|{\cal M}_{2\nu}|^{exp}$      &  $   <0.081              $
                             &  $   0.280_{-0.010}^{+0.006}$
                             &  $   0.141_{-0.014}^{+0.004}$
                             &  $   0.294_{-0.033}^{+0.029}$
                             &  $   0.038_{-0.01}^{+0.01} $
                             &  $   0.027_{-0.01}^{+0.01} $\\
${\cal M}_{2\nu}^{cal}$ &0.091&0.100&0.121&0.102&0.118&0.096\\
\hline\hline
\end{tabular}
\label{tab2}
\caption{ \protect \footnotesize
Upper bounds on the effective neutrino mass $<m_{\nu}>$ (in eV)
 obtained from the QRPA calculations of the nuclear matrix elements. For the
sake of comparison, in all the cases the same experimental data, as well
the same effective axial vector coupling
constant ($g_A=-g_V$) have been used.}
\vspace{3mm}
\begin{tabular}{lcccccccc}
\hline     \hline
&&{$^{48}Ca$}& {$^{76}Ge$}& {$^{82}Se$}& {$^{100}Mo$}&
{$^{128}Te$}& {$^{130}Te$}\\
\hline
Pasadena &(ref. \cite{pas})                      &&$4.4$&$20$&$20$&$1.8$&$22$\\
Heidelberg& (ref. \cite{hei})                       &22&2.0&7.4&26&1.5&21\\
T\"{u}bingen& (ref. \cite{tub})                  &&$3.1$&$12$&&$3.8$&$31$\\
our results &(ref. \cite{Krm94})
&$71$&$1.5$&$5.3$&$8.8$&$1.0$&$12$\\
\hline\hline
\label{tab3}
\end{tabular} \end{center} \end{table} 
\begin{thebibliography}{99}

\bibitem{Reviews} See W.C. Haxton and G.J. Stephenson, Prog. Part. Nucl. Phys.
{\bf 12} (1984) 409;
M. Doi, T. Kotani and E. Tagasuki, Prog. Theor. Phys. Suppl. {\bf 83} (1985) 1;
F.T. Avignone III and R.L. Brodzinski, in {\it Neutrinos}, edited by H.V.
Klapdor (Springer-Verlag, New-York, 1988), p. 147; K. Muto and H.V. Klapdor,
{\it ibid.} p. 183;
F. Krmpoti\'{c}, in {\it Lectures  on  Hadron Physics},
Ed.  by  E. Ferreira (World Scientific, Singapore, 1990) p. 205
\bibitem{Tom91} T. Tomoda, Rep. Prog. Phys. {\bf 54} (1991) 53, and references
therein
\bibitem{Kra87} J. Kramp, D. Habs, R. Kroth, M. Music, J. Schirmer, D. Schwalm
and C. Broude, Nuc.\ Phys.\ {\bf A474} (1987) 412
\bibitem{Hir90} J Hirsch and F. Krmpoti\'{c}, Phys. Rev. {\bf C41} (1990) 792;
J. Hirsch, E. Bauer and F. Krmpoti\'{c}, Nucl. Phys. {\bf A516} (1990) 304;
J. Hirsch and F. Krmpoti\'{c}, Phys. Lett. {\bf B246} (1990) 5
\bibitem{Krm92} F. Krmpoti\'{c}, J. Hirsch and H. Dias,
Nucl. Phys. {\bf A542} (1992) 85
\bibitem{Krm93} F. Krmpoti\'{c} Phys. Rev. {\bf C48} (1993) 1452
\bibitem{Krm94} F. Krmpoti\'{c} and S. Shelly Sharma,
Nucl. Phys. {\bf A} (to be published)
\bibitem{Ala71} G. Alaga, F. Krmpoti\'{c}, V. Lopac, V. Paar and L. \v{S}ips,
Fizika {\bf 4} (1971) 25
\bibitem{Krm93a} F. Krmpoti\'{c}, A. Mariano, T.T.S. Kuo and K. Nakayama,
Phys. Lett. {\bf B319} (1993) 393
\bibitem{Gri92} A. Griffiths and P. Vogel, Phys. Rev. {\bf C46} (1992) 181
\bibitem{Sey91} H. Seyfarth et al. Z. Phys. {\bf A339} (1991) 269
\bibitem{Nak82}  K. Nakayama, A. Pio Gale\~{a}o  and  F. Krmpoti\'{c},
Phys. Lett. {\bf B114} (1982) 217
\bibitem{pas}
J. Engel, P. Vogel and M.R. Zirnbauer, Phys. Rev. {\bf C37}, 731 (1988).
\bibitem{hei}
K. Muto and H.V. Klapdor, Phys. Lett. {\bf 201B}, 420 (1988);
K. Muto, E. Bender and H.V. Klapdor-Kleingrothaus,
Z. Phys. {\bf A339}, 435 (1991).
\bibitem{tub}
J. Suhonen, S.B. Khadkikar and A. Faessler,
Nucl. Phys. {\bf A535}, 509 (1991).
\end{thebibliography}
\end{document}